\documentclass[%
 reprint,
superscriptaddress,
 amsmath,amssymb,
 aps,
 prb,
]{revtex4-2}
\usepackage[colorlinks=true,urlcolor=blue,citecolor=blue]{hyperref}
\usepackage{txfonts}
\usepackage{xcolor}
\usepackage{MnSymbol}
\usepackage{scalerel}

\usepackage{graphicx}
\usepackage{dcolumn}
\usepackage{bm}

\newcommand{\bigfilleddiamond}{\raisebox{-0.1\dimexpr\ht\strutbox}{\scalebox{1.5}{$\filleddiamond$}}}

\begin{document}

\preprint{APS/123-QED}

\title{Control of Ferrimagnetic Compensation and Perpendicular Anisotropy in Tb$_x$Co$_{(100-x)}$ with H$^{+}$ ion implantation}

\author{Robbie G. Hunt}
 \affiliation{%
 Uppsala University, Department of Physics and Astronomy 
}%
\author{Dmitrii Moldarev}%
\affiliation{%
  Uppsala University, Department of Physics and Astronomy 
}%
\author{Matías P. Grassi}%
\affiliation{%
  Uppsala University, Department of Physics and Astronomy 
}%

\author{Daniel Primetzhofer}%
\affiliation{%
  Uppsala University, Department of Physics and Astronomy 
}%

\author{Gabriella Andersson}
 \affiliation{%
 Uppsala University, Department of Physics and Astronomy 
}%


\begin{abstract}

The tuning of magnetic properties through electrochemical loading of hydrogen has recently attracted significant interest as a way to manipulate magnetic devices with electric fields. In this paper we investigate quantitatively the magneto-ionic effect of hydrogen uptake on the magnetic properties of rare-earth transition metal alloy Tb$_x$Co$_{(100-x)}$ in the composition range of $x=10-39$ at.\% using ion implantation. Using this technique we are able to link changes in magnetic behaviour to exact concentrations of hydrogen, isolated from the movement of any other ions that would be a factor in electrochemical studies. The composition of the alloy has been varied alongside the hydrogen dose to characterize the effect of progressive hydrogen loading on the full range of $x$ displaying out-of-plane magnetic anisotropy. We find large changes in two important properties: the compensation composition and the Co-rich in-plane to out-of-plane magnetic anisotropy transition composition, both of which move by 6 at.\% towards higher Tb concentrations after hydrogen implantation. This shift in composition does not increase with a larger dose. From the changes in magnetization we attribute the change in compensation composition to a significant reduction of the moment on the Tb sublattice.

\end{abstract}

\maketitle


\section{Introduction}
The development of magnetoelectric materials has been of significant interest for the development of new devices, due to the promise of greater energy efficiency and unique functionalities \cite{manipatruni2019scalable, jaiswal2017proposal}. Few single-phase magnetoelectric materials exist, such as the room-temperature multiferroic BiFeO$_3$ \cite{ederer2005weak, heron2014electric}, and so a great deal of work is dedicated to indirect means of achieving magnetoelectric coupling. These indirect approaches include devices based on interfacial strain, both static \cite{franke2015reversible,hunt2023strain} and more recently dynamic \cite{camara2019field, shuai2022local} variations, modification of the carrier densities in ferromagnetic semiconductors \cite{ohno2000electric}, and manipulation via ionic motion\cite{navarro2018large}.

Magneto-ionic devices have recently attracted significant interest in the field of spintronics as a way to control magnetic properties through application of a voltage, including perpendicular magnetic anisotropy \cite{bhatnagar2023controlling}, Ruderman–Kittel–Kasuya–Yosida interlayer exchange \cite{tran2024field}, Dzyaloshinskii-Moriya interaction \cite{herrera2019nonvolatile}, and exchange bias \cite{hasan2023large}. These properties are all extremely relevant to device applications but are otherwise difficult to reversibly manipulate post-fabrication. 

A significant downside to magneto-ionic devices is the slow response time of ionic elements. Typical charging times for oxygen-based devices are on the order of 15 minutes \cite{de2022voltage}, 4 minutes for systems making use of nitrogen \cite{de2020voltage}, and a few seconds for Li-based devices \cite{ameziane2022lithium}. For oxygen and nitrogen based magneto-ionics much of the work done for these makes use of ionic liquids or liquid electrolytes to amplify electric fields or source ions, and while this is sufficient for a proof of concept experiment it is not practical for industrial application. Lithium has the enormous advantage of well-studied solid-state electrolytes, and has also been used on much thicker in-plane films from which it can be understood that the Li ions have excellent diffusivity in relevant materials \cite{pravarthana2019reversible}. However, the lifespan of lithium-electrolytes and battery technology is typically on the order of 500-2000 cycles before electrolyte degradation becomes an issue \cite{zhang2022high}, whereas the cyclability requirements for computational applications can be many orders of magnitude higher than this \cite{takemura201032}. 

In these respects, hydrogen-based systems presents a good alternative candidate ion system due to the reported charging times of only a few tens of microseconds, the development of solid-state electrolytes, the high diffusivity of hydrogen and the large proven cyclability of devices \cite{huang2021voltage}. 

Previous work in the literature using hydrogen as an ionic species has found significant success in the manipulation of both a Pt/Co system exhibiting perpendicular magnetic anisotropy (PMA) \cite{tan2019magneto}, the modulation of inter-layer RKKY coupling \cite{kossak2023voltage}, and the compensation behaviour of rare-earth transition-metal (RE-TM) alloy GdCo \cite{huang2021voltage} - demonstrating the wide range of applicability. In the case of GdCo, there are significant changes in the magnetic sublattice coupling leading to a large reduction of the compensation temperature upon hydrogen loading, allowing for effective control over the net magnetization direction. Recently, in further work using an electrolyte source of hydrogen ions, it was found that for Tb$_x$Co$_{(100-x)
}$ films with compensation temperatures below room temperature it is possible to manipulate the PMA strength to induce a voltage-controlled in-plane to out-of-plane transition \cite{xiao2023hydrogen}. 

However, in electrochemical systems there can be more than one ionic species that moves in response to an electric field - including passivated oxygen ions. The polarized neutron reflectometry results of Sheffels et al. \cite{sheffels2023insight} on a Pt/Co/GdOx system revealed some of the complexities of trying to understand the voltage-driven effects, with a complicated mix of oxide and hydroxide phases contributing to the reduction of PMA strength. The combination of multiple ionic species makes it difficult to disentangle the effects of each contributing species which in turn makes it challenging to accurately model and predict future behaviour through methods such as density functional theory. To simplify the system, it is advantageous to consider a non-chemical loading mechanism  such as ion implantation to study the effects of hydrogen loading on the magnetic behaviour of a magnetic material.

\begin{figure*}
    \centering
    \includegraphics[width=\linewidth]{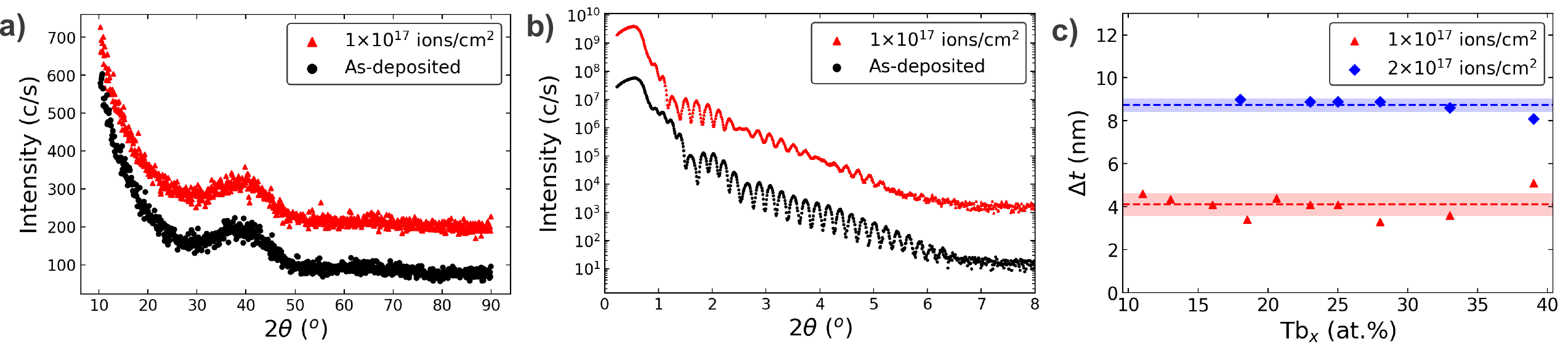}
    \caption{Changes in the a) GIXRD and b) XRR patterns for a Tb$_{x}$Co$_{(100-x)}$ sample after an implanted hydrogen dose of 1$x10^{17}$ ions/cm$^2$. Intensities for the implanted sample are offset for clarity. c) Thickness expansion of the Tb$_x$Co$_{(100-x)}$ layer for varying composition based on XRR fits before and after implantation. Dashed lines show the average thickness expansion for each dose, and shaded regions represent one standard deviation.}
    \label{fig:gixrd-xrr}
\end{figure*}

In this work we investigate the hydrogen-induced changes in the magnetic properties of Tb$_x$Co$_{(100-x)}$ amorphous thin film alloys by ion implantation. By using this technique we isolate the magneto-ionic effects to hydrogen ions alone, where electrochemical means can have additional reactions involving hydroxide and oxide phases either forming or migrating in response to the applied electric field. Like GdCo alloys, this material is ferrimagnetic and exhibits both a compensation composition at room temperature a bulk-like PMA with two transition compositions from in-plane anisotropy (IPA) to PMA with one in the Co-rich regime (which we will refer to as $x_T$) and one in the Tb-rich region. Using ion beam analysis techniques we obtain precise knowledge of the exact alloy composition and hydrogen concentration to investigate the change in PMA strength, compensation behaviour, and saturation magnetization. We investigate these changes in response to two implanted doses of hydrogen ions. We put forward an explanation for the reduction in Tb atomic magnetic moment based on the sperimagnetic nature of these alloys, assuming that the Tb cone is altered by the introduction of hydrogen and that the cone angle is increased.

\section{Sample growth}

 Samples are grown using a DC magnetron sputtering technique in a UHV chamber with a base pressure of $6\times 10^{-10}$ Torr or better. The thin films are deposited at an Ar pressure of 2.06 mTorr onto silicon substrates with a native oxide layer. Prior to deposition substrates are degassed at 200 \textdegree C for 40 minutes to clean the substrate surface. The TbCo layer is grown directly onto the substrate by co-deposition, with the Co power varied to obtain the desired composition and an Al layer is used as a cap to protect against oxidation. 

The full sample structure for all films is Si / Tb$_x$Co$_{(100-x)}$ (30 nm) / Al (8 nm), with a standard deviation of $\pm2$ nm for the TbCo thickness across all compositions. From previous studies in the literature \cite{ciuciulkaite2020magnetic}, this degree of variation in the thickness is unlikely to affect the magnetic properties as above 20 nm the coercivity, a key indicator of the coupling strength and anisotropy, remains stable up to 100 nm.

Film quality and structure are investigated through X-ray measurements using a Cu K-$\alpha$ source with a wavelength of $\lambda=1.54$ Å.The amorphous character of the entire composition range is confirmed through grazing incidence X-ray diffraction (GIXRD) with an incident angle of $\omega=0.5$$^{\circ}$.  For samples across the entire composition range of interest there is a broad amorphous peak but no crystalline peaks, shown in Fig. \ref{fig:gixrd-xrr}a), and so the samples are considered X-ray amorphous.

Layer thicknesses are quantified by fitting X-ray reflectometry (XRR) data using GenX \cite{bjorck2007genx} and the Tb concentration, $x$, of the Tb$_x$Co$_{(100-x)}$ alloys is verified through Rutherford backscattering spectrometry (RBS) using 2 MeV He as the primary beam and the data is fit using SIMNRA \cite{mayer1999simnra}. An indicative XRR profile is shown in Fig. \ref{fig:gixrd-xrr}b) and demonstrates that the films are smooth with Kiessig fringes up to $2\theta$=7$^{\circ}$.

\section{Hydrogen implantation}

Samples are implanted with 2.5 keV hydrogen at the 350 kV Danfysik Implanter at the Tandem laboratory of Uppsala University \cite{strom2022ion}. The beam is raster scanned across all samples on the plate to obtain a homogeneous distribution. On each sample plate a series of Tb$_x$Co$_{(100-x)}$ samples with varying compositions are included such that the dose received by each sample is the same. The implanted dose is measured using a set of Faraday cups placed around the samples.

\begin{figure}
    \centering
    \includegraphics[width=\linewidth]{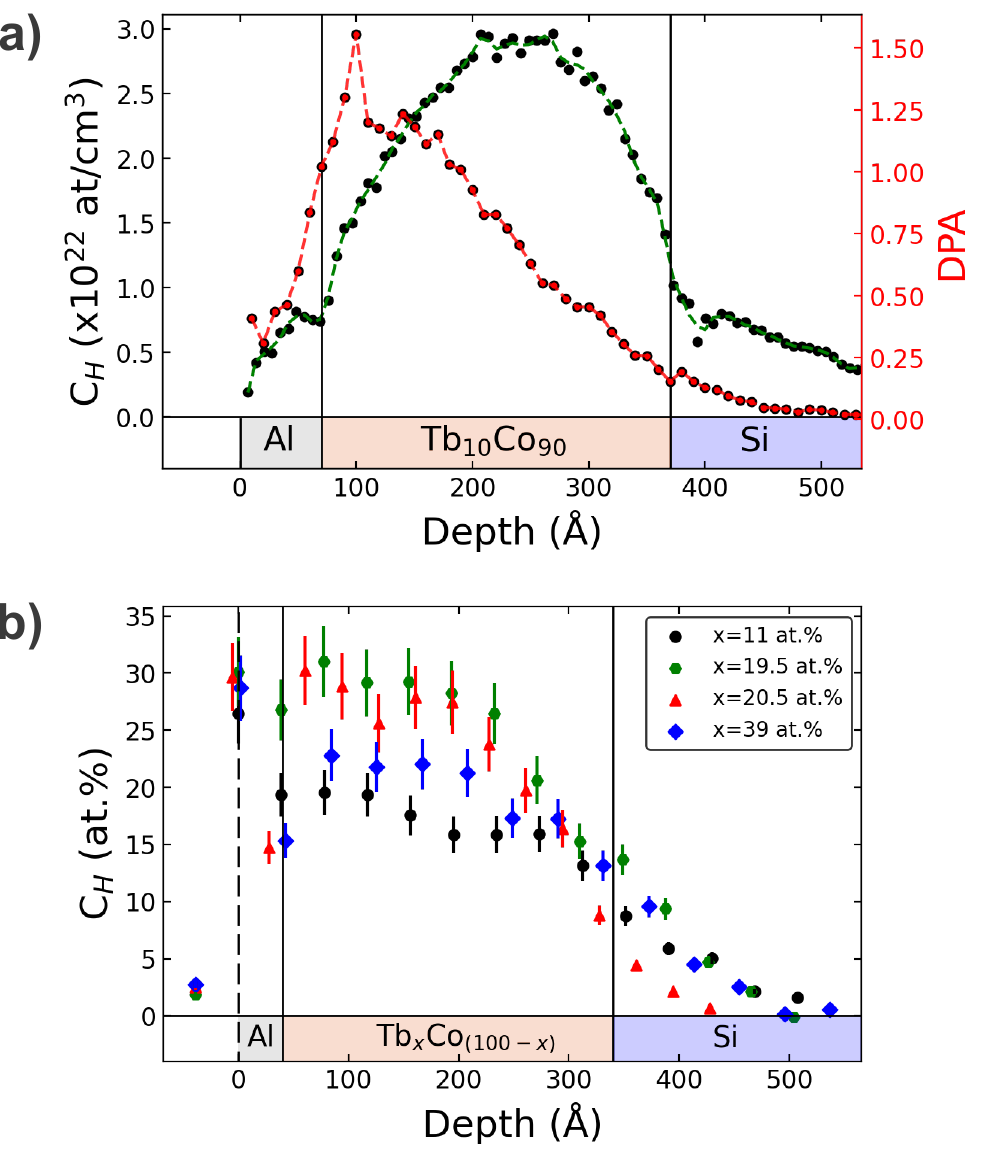}
    \caption{a) Simulated implanted hydrogen ion distribution profile and imlantation damage profile in a sample structure of Al/Tb$_{10}$Co$_{90}$/Si for an ion energy of 2.5 keV and dose of 1$\times$10$^17$ ions/cm$^2$. b) Measured hydrogen profile from NRA of a Tb$_{10}$Co$_{90}$ sample after implantation of 1$\times10^{17}$ ions.}
    \label{fig:srim-nra}
\end{figure}

Figure \ref{fig:srim-nra}a) shows the expected implantation profile for H$^{+}$ ions with an energy of 2.5 keV simulated using the SRIM software \cite{ziegler2010srim}. Prominent in this graph is a peak in the centre of the TbCo film, with some hydrogen trapped in the capping layer and a trailing edge of hydrogen atoms leading into the silicon substrate. Since the distribution is well centered within the film, this energy was chosen as the implantation energy. It is also important that we take into account the damage done by the irradiation process. In a recent study by Krupinski et al. \cite{krupinski2021control} investigating irradiation of RE-TM films by He and Ne species, significant changes in magnetic behaviour were obtained purely by ion-induced damage, calculated as the displacement per atom (DPA) caused by the irradiation process. From our simulations we are also able to calculate the damage profile in Fig. \ref{fig:srim-nra}a) and we find that our profile exhibits a peak towards the surface of the film, where the profiles investigated by Krupinski et al. have a peak in the middle of the film.

We consider two nominal doses for implantation: $1\times10^{17}$ ions/cm$^2$ and $2\times10^{17}$ ions/cm$^2$ of hydrogen atoms, which corresponds to a ratio of approximately 1:2.5 ($\approx$28.5 at.\%)  and and 2:2.5 ($\approx$44 at.\%) hydrogen atoms per metal atom based on the areal densities obtained from RBS. 

The potential structural changes caused by implantation are probed by x-ray reflectometry and GIXRD after implantation. From the GIXRD (Fig. \ref{fig:gixrd-xrr}a) we see no evidence of a change in the amorphous peak or emergence of any crystalline phases resulting from the implantation. From the reflectometry in Fig. \ref{fig:gixrd-xrr}b), we see that the H implantation does introduce changes in electronic density, thickness and surface roughness (from approximately 5Å to 10Å in the surface layer) which  altogether leads to a decrease in intensity of the Kiessig fringes in the reflectometry profile. As the origin of the PMA in these films are bulk-like and the films are sufficiently thick we do not expect either the change in roughness or thickness to significantly impact the magnetic properties of the film. Notably, the absolute thickness expansion, $\Delta t$, in the Tb$_x$Co$_{(100-x)}$ layers after the implantation is consistent across the composition range for both doses and is found to be (4.1$\pm$0.5) nm for a dose of 1x$10^{17}$ ions/cm$^2$  and (8.7$\pm$0.3) nm for a dose of 2x$10^{17}$ ions/cm$^2$. The consistency in the layer expansion provides a good initial indicator that the implantation process is uniform for all of the samples as this expansion is consistent for all compositions. The origin of this thickness expansion may be a combination of hydrogen-induced lattice expansion \cite{bylin2022hydrogen} and radiation-induced swelling \cite{swenson2018swelling}, but it is difficult to examine the exact nature of the thickness increase. Disentangling these contributions goes beyond the scope of this study and we use this information merely to show that the implantation is consistent for all samples.

Although the implantation dose is the same, we can expect that the final hydrogen concentration of each sample may vary as a result of many factors. These factors include changes in hydrogen solubility, differences in atomic density relating to the change in Tb$_x$Co$_{100-x}$ composition, and small variations in sample quality during the growth process that will affect hydrogen retention such as the film thickness and porosity. As such an essential element of this study is to measure the retained concentration of hydrogen.

After implantation two ion beam based techniques are used to measure the hydrogen content: nuclear reaction analysis (NRA) which is used to measure hydrogen depth profile with a depth resolution of a few nm, and elastic recoil detection analysis (ERDA) to measure the H concentration in the film and evaluate the uniformity of the implantation process across all samples. NRA was conducted using $^{15}$N beam with \textgreater 6.385 MeV, and the procedure is described in more detail elsewhere \cite{komander2019hydrogen}. A primary beam of 2.1 MeV He$^{+}$ ions was employed for ERDA measurements. Recoils were detected with a solid-state detector positioned at the forward scattering angle of 40°, while the samples were rotated 70° with respect to the beam. For more experimental details we refer to Ref. \cite{pitthan2023thin}. The hydrogen concentration has been fit by assuming the majority of the ions are inside the Tb$_x$Co$_{(100-x)}$ layer, as suggested by the NRA results.

The obtained hydrogen concentration profiles from NRA in Fig. \ref{fig:srim-nra}b) show a flatter distribution of hydrogen atoms than expected from SRIM simulations with a small bias towards the surface of the film for the fluence of $1\times10^{17}$ ions/cm$^2$, indicating that the hydrogen is able to diffuse through the film as compared to the calculations of Fig. \ref{fig:srim-nra}a), where there is no diffusion. This diffusivity may be because the ions are not strongly bound to the initial implantation sites. Moreover, the profile might be biased by the irradiation-induced damage, but not solely determined by it. Regardless, this creates a much more homogeneous hydrogen profile in which the concentration through the film thickness can be considered approximately equal.

Equally of note is that the hydrogen concentrations vary somewhat between the measured samples. Samples with the highest and lowest Tb concentrations show the smallest retained hydrogen concentrations, while the samples around $x=20$ have a much higher hydrogen concentration, approximately 30 at.\% compared to 20 at.\%.
 
\begin{figure}
    \centering
    \includegraphics[width=\linewidth]{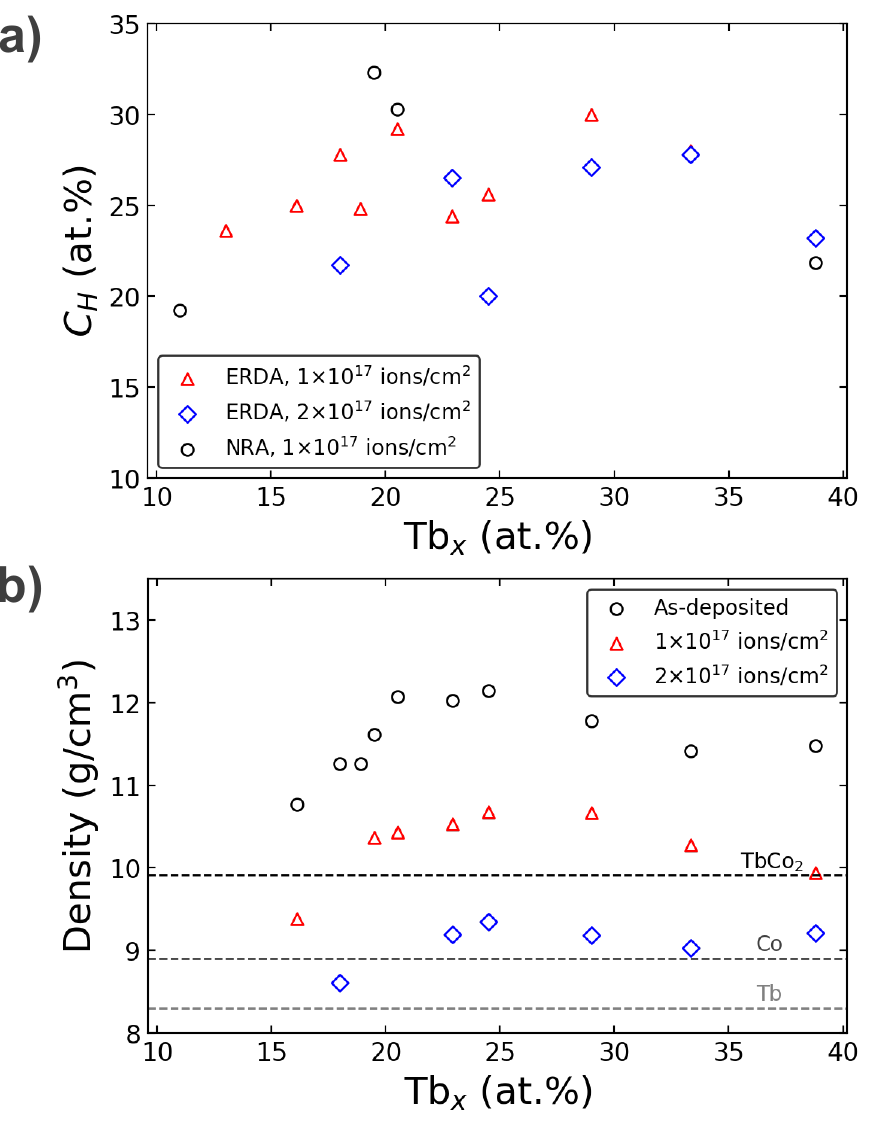}
    \caption{a) Composition dependence of hydrogen concentration, as a percentage relative to the metal atoms, determined from ERDA and NRA for implantation dosse of 1x10$^{17}$ and 2x10$^{17}$ ions/cm$^2$.  b) Mass density derived from RBS and XRR measurements for the as-deposited and implanted samples. Densities for implanted samples use the increased film thickness measured in Fig. \ref{fig:gixrd-xrr}c). }
    \label{fig:erda}
\end{figure}

To examine this in more detail we perform ERDA on all samples from both doses. Fig \ref{fig:erda} shows that there is no strong dependence of the retained dose on either the Tb concentration or the implanted hydrogen dose. For the lower dose, there may be a small linear trend with Tb concentration but this is dominated much more by the scatter of the data around the average value of 26.5 at.\%. The films with highest (39 at.\%) and lowest (11 at.\%) terbium concentrations retain the lowest concentration of hydrogen which suggests that at these doses the concentration of the RE element is not contributing significantly to the retention.

From the areal density (at./cm$^2$) obtained from RBS data and the thickness obtained from XRR data we are able to calculate the atomic density of the film which is then converted to the mass density by weighing the mass based on the composition,

\begin{equation}
    \rho = \frac{1}{N_A} \left[ \frac{(100-x)}{100}\rho_{at.} \times \rho_{Co} + \frac{x}{100}\rho_{at.} \times \rho_{Tb})\right],
\end{equation}

where $N_A$ is Avogadro's constant, $\rho$ is the density in units of g/cm$^3$, $\rho_{at}$ is the density in units of at./cm$^3$, and $\rho_{Co}$ and $\rho_{Tb}$ are the mass densities of Co and Tb respectively. The as-deposited films have densities that are significantly higher than the densities of either atomic species or the intermetallic TbCo$_2$ compound. It is unclear why this is as for thicker 100 nm films (not shown here) the calculated mass densities using a similar approach are closer to the densities of the constituent elements with $\approx$8.5 g/cm$^3$. When calculated taking the thickness expansion into account after ion implantation, the densities then converge towards mass densities that are more similar to that of Co. The origin of this discrepancy is unclear, but what we can understand from this is that there is a Tb-composition dependence of the mass density that exhibits a maximum in the region of $x\approx24$ at.\% at room temperature. 

The relationship here between density and hydrogen solubility may then be the opposite of expected, with a general trend of the hydrogen concentration increasing with the density (as opposed to decreasing due to the increased number of atoms). However this relationship is weak and the most critical conclusion we can draw from these measurements is that the hydrogen concentration is at saturation for these films.

\begin{figure*}
    \centering
    \includegraphics[width=1.4\columnwidth]{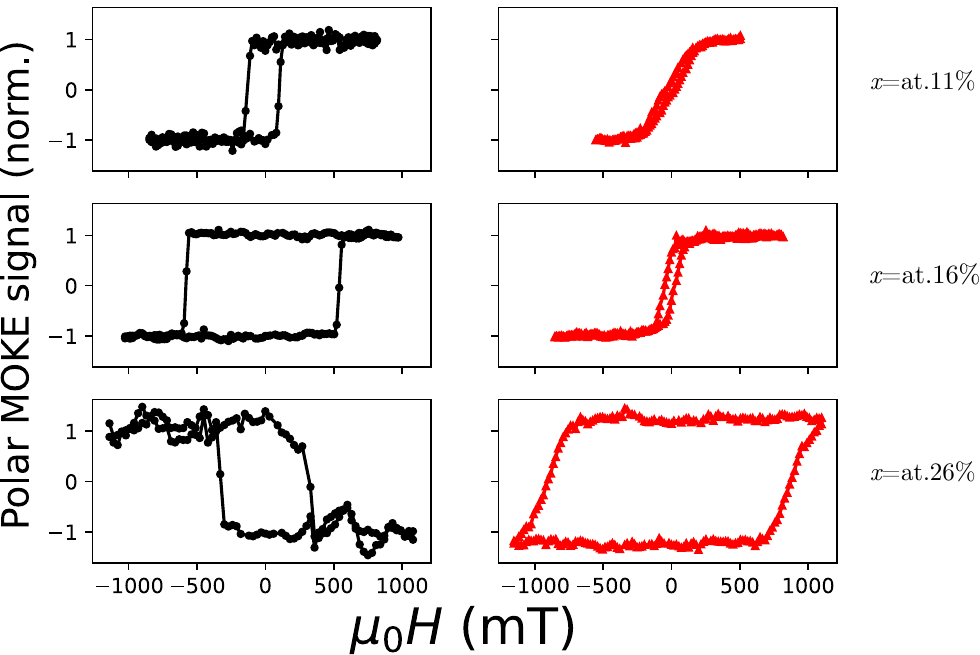}
    \caption{MOKE hysteresis loops for as-deposited (black) and implanted (red) samples with a nominal dose of $1\times10^{17}$ ions/cm$^2$. The change in polarity from x=16 at.\% to x=26 at.\% corresponds to crossing the compensation composition.}
    \label{fig:hysteresis_loops}
\end{figure*}

\section{Magnetic Characterization}

Magnetic hysteresis loops at room temperature are performed using a magneto-optic Kerr effect (MOKE) experiment in the polar geometry with a laser wavelength of 670 nm. From previous work in the literature \cite{khorsand2013element}, it is known that the Kerr effect for these ferrimagnetic materials is not sensitive purely to the net magnetization but is instead more sensitive to the magnetization on one particular sublattice dependent upon the wavelength of the probing laser. At this wavelength, the Kerr effect is in principle more sensitive to the Co sublattice. The experimental setup is such that the angle between the polarizer and the analyzer remains fixed and so the change in orientation of the Co sublattice across the compensation composition is clear. The inversion of the hysteresis polarity is used to help define the position of the room temperature compensation composition, $x_c$, as shown in Fig. \ref{fig:hysteresis_loops} together with the increase in coercivity. Prior to implantation, all samples have square hysteresis loops suggesting excellent uniaxial PMA with no significant in-plane contributions to the magnetic anisotropy. Since all of the samples display PMA we take the position of the Co-rich in-plane anisotropy to out-of-plane transition, $x_T$, in the as-deposited state to be $\approx$10 at.\%, based on work reported in the literature for similar samples \cite{odagiri2024coexistence, thorarinsdottir2023competing}.

\begin{figure}
    \centering
    \includegraphics[width=\columnwidth]{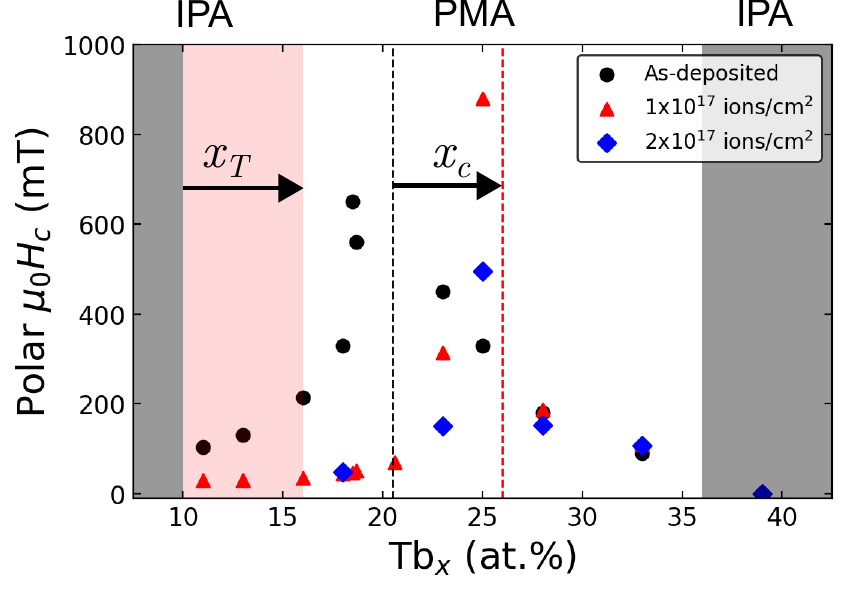}
    \caption{Determination of compensation composition, $x_c$, in the as-deposited state ($\medbullet$) and after hydrogen implantation doses of $1\times 10^{17}$ ions/cm$^2$ ($\color{red}\filledmedtriangleup$) and $2\times 10^{17}$ ions/cm$^2$ ($\color{blue}\bigfilleddiamond $). Vertical lines indicate the approximate position of $x_C$, with the red dashed line showing the position for both implantation doses. Shaded regions indicate compositions for which the films have in-plane anisotropy, with the left-hand side region marked as $x_T$, and the lighter red region being the change in the IPA region after an implantation dose of $1\times 10^{17}$ ions/cm$^2$.}
    \label{fig:implant_comp}
\end{figure}

In agreement with previous results \cite{yagmur2021magnetization, ciuciulkaite2020magnetic, alebrand2012light} for Tb$_x$Co$_{(100-x)}$ alloys, the value of $x_c$, at which point the Tb and Co sublattice magnetizations perfectly compensate each other, is found to be approximately 20.5 at.\% in Fig. \ref{fig:implant_comp}, indicated by the point at which the coercivity of the film diverges and is accompanied by a reversal of hysteresis loop polarity as explained previously. A sample grown with this composition had a coercive field in excess of 1.2 T and it was not possible to measure the hysteresis behaviour of the sample, in line with what is expected at or extremely near to compensation and so we take this composition to be the value of $x_c$ prior to implantation. 

\begin{figure}
    \centering
    \includegraphics[width=0.8\columnwidth]{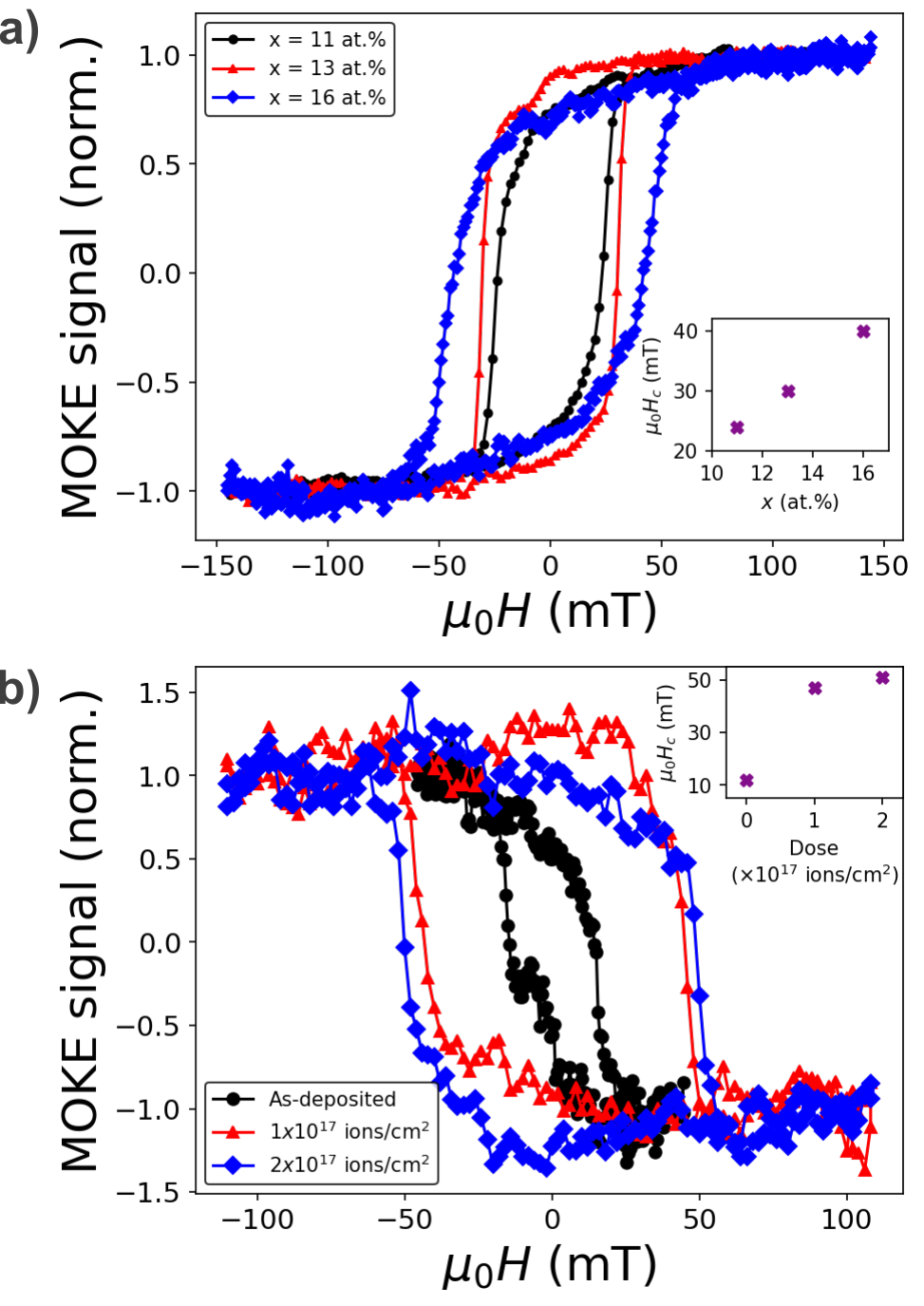}
    \caption{In-plane magnetic hysteresis loops for a) $x = 10 - 16$ at.\% after a $1\times10^{17}$ ions/cm$^2$ dose and b) $x = 39$ at.\% under varying hydrogen dose. Insets show the coercivities from each hysteresis loop.}
    \label{fig:ip_loops}
\end{figure}

With the implanted hydrogen dose of $1\times10^{17}$ ions/cm$^2$, there is an immediate shift of the compensation composition towards a higher value of $x$ by approximately 5-6 at.\%. In addition to this, the value of $x_T$ increases by approximately the same magnitude with the $x= 11, 13$ at.\% samples falling in-plane after implantation and the $x=16, 18$ at.\% samples displaying some tilted anisotropy with in-plane and out-of-plane components. The reduced PMA strength in these samples is likely due to the implanted hydrogen disrupting the ratio of in-plane and out-of-plane Tb-Co, Tb-Tb and Co-Co pair correlations that are linked to the PMA in these materials \cite{harris1992structural}. For the samples with IPA, we show the hysteresis behaviour measured in a longitudinal MOKE configuration in Fig. \ref{fig:ip_loops}a). For the samples that lose PMA as a result of implantation, the coercivity increases linearly with the value of the Tb concentration.

We can initially understand the change in compensation composition as a change in moment on the terbium atoms. From the conditions for compensation, we know that the overall moment on the terbium atoms must be lower upon introducing hydrogen and so a greater amount of terbium atoms is needed to reach a perfect balance between the two sublattices. This explanation agrees well with the explanation proposed by Huang et al. \cite{huang2021voltage} where a reduction in compensation temperature of a GdCo layer was attributed to the reduction in moment of the Gd and Co species upon hydrogen loading driven by the structural changes in the Gd-Co bond lengths and the resulting reduction in the exchange coupling constants.

We now consider samples implantated with a dose of $2\times10^{17}$ ions/cm$^2$. Here samples have been implanted on a more constrained range of $x=17 - 39$ at.\%, constrained as samples below $x=17$ at.\% are already expected to lose PMA. The lower bound was chosen based on the shift in $x_T$ for a dose of $1\times10^{17}$ ions/cm$^2$. From the previously presented ERDA data in Fig. \ref{fig:erda} we understand that there is no significant increase in hydrogen concentration as compared with the initial dose, but from the increased dose we can expect a higher degree of ion-induced damage. Despite this, we see no significant change in the positions of either $x_T$ or $x_c$, indicating that the magnetic properties are not sensitive to the damage induced in implantation for hydrogen at these energies or that there is no significant damage to the amorphous structure. Turning to the behaviour of the $x=39$ at.\% samples that always have in-plane anisotropy, Fig. \ref{fig:ip_loops}b) shows that there is a minimal increase in coercivity when the dose is doubled. From the previously presented data on the thickness expansion in Fig. \ref{fig:gixrd-xrr}c) we can understand that the thickness of the sample continues to increase significantly with this increased dose and so we can understand that while there is some minor variation in the magnetic properties, the thickness expansion and the ion-induced damage are not the leading causes of the changes in $x_c$ and $x_T$.

\begin{figure*}
    \centering
    \includegraphics[width=2\columnwidth]{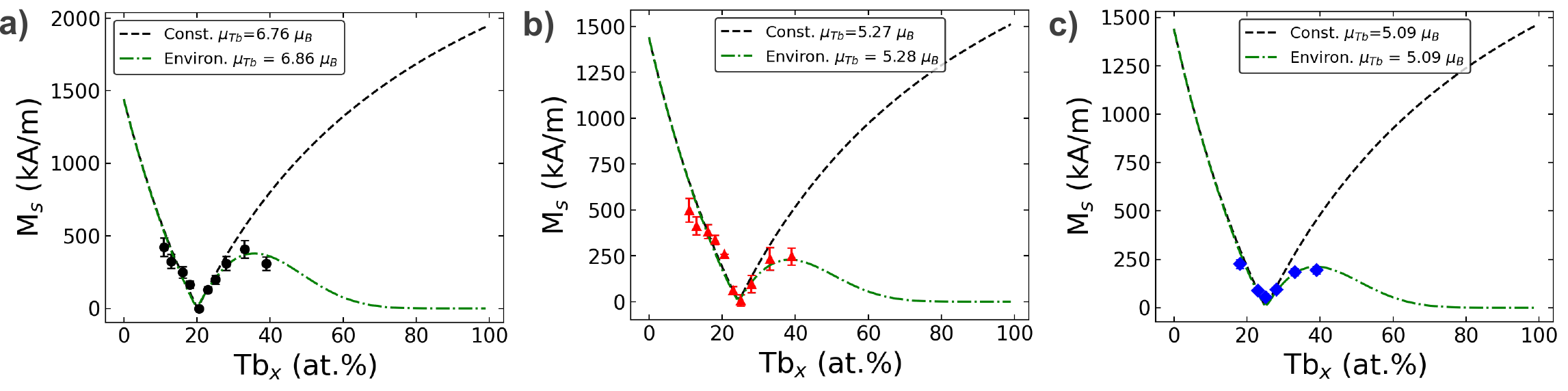}
    \caption{Saturation magnetization as a function of composition for the a) as-deposited, b) $1\times10^{17}$ ions/cm$^2$ and c) $2\times10^{17}$ ions/cm$^2$ H$^{+}$ implantation doses. Black and green dashed lines represent best fits to a constant magnetization and environmental-dependent magnetization model, respectively (see text for more detail). } 
    \label{fig:ms_comp}
\end{figure*}

To examine this in more detail we perform magnetometry to extract the net saturation magnetization of the ferrimagnetic moment as a function of composition shown in Fig. \ref{fig:ms_comp}. The composition-dependent curve can be related to the atomic magnetic moments of the Co and Tb sublattices through two models as shown previously by Suzuki et al. \cite{suzuki2023thickness}. First, through an initial naive model in which the atomic magnetic moment of the Co and Tb atoms remains constant for all compositions. In this model the net saturation magnetization is calculated as,

\begin{equation}
    M_s(x) = \lvert \frac{m_{z,Co}(100-x) - m_{z,Tb}x}{v_{Co}(100-x) + v_{Tb}x}\rvert,
\end{equation}

where $M_s$ is the net saturation magnetization of the ferrimagnetic moment, $m_{z,Co}$ and $m_{z,Tb}$ are the atomic Co and Tb magnetic moments in the $z$ direction, and $v_{Co}$ and $v_{Tb}$ are the atomic volumes of Co and Tb taken to be the same as in Ref. \cite{suzuki2023thickness}, $v_{Co} = 1.10\times10^{-29} m^3$ and $v_{Tb} = 3.211\times10^{-29} m^3$. This model fits the compensation behaviour reasonably well, but cannot fit the decrease in $M_s$ at higher values of $x$ far from compensation.

To encapsulate this decrease in magnetization we use a modified model in which the atomic magnetic moments vary with increasing Tb composition, used initially by Jaccarino and Walker \cite{jaccarino1965discontinuous} to describe magnetization of impurity species in alloy materials. The composition dependence of the atomic moment is then described by,

\begin{equation}
    m_{z, i}(x) = P_j(x)m_{z, i}(x=0),
\end{equation}

where the $i$ is the atomic species (Co or Tb), and $P_j(x)$ is a probabilistic prefactor accounting for the change in atomic environment with increasing $x$,

\begin{equation}
    P_j(x) = \sum\limits_{k=j}^{N} \binom{N}{k} (100-x)^kx^{N-k},
\end{equation}

with $j$ being the minimum number of nearest Co neighbours, $N$ the coordination number which for TbCo takes a maximum of $N=12$, and $\binom{N}{k}$ is the binomial coefficient.

For each dose we show in Fig. \ref{fig:ms_comp} fits to both of these models. To simplify the fitting, we make the assumption that the magnetic moment of Co at $x=0$ is always the expected atomic magnetic moment of 1.71$\mu_B$. For the as-deposited samples this justification is simple as the saturation magnetization at $x=0$ should converge to the atomic moment of Co. In the implanted samples, we turn to the x-ray magnetic circular dichroism data of Huang et al. \cite{huang2021voltage} obtained for GdCo in which the change in Co atomic moment upon hydrogen loading is observed to be extremely small with differences only becoming discernible when fitted models were extrapolated out to much higher temperatures than room temperature. Given also that the value of $x_c$ shifts towards higher Tb concentrations it is also reasonable to assume that the largest changes in magnetic moment occur on the Tb sublattice as opposed to the Co sublattice. 

Under these assumptions, the change in $x_c$ can be understood to be result from a change in the extrapolated value of $m_{Tb}(x=0)$ and that the implanted hydrogen ions behave consistently across all samples regardless of change in rare-earth concentration. In previous work in the literature, it has been concluded in other RE-TM alloys that these changes could be explained by the intercalation of hydrogen leading to larger RE-TM bond lengths which  reduces the magnitude of the exchange coupling \cite{huang2021voltage}. In the ab initio modelling presented in Ref. \cite{huang2021voltage} for GdCo, hydrogen concentrations of 20 and 40 at.\% were considered in a crystalline GdCo$_2$ system and predict a continuous degradation of the exchange coupling as the concentration increases. Based on our NRA and ERDA results, we consider that all samples are saturated with hydrogen although the exact value of concentration varies possibly due to the change in density. Between the two doses, the most obvious change will be the increase in DPA which we expect to scale linearly with the dose. Despite this, we find that fitting to either model shows similar reductions in the magnetic moment on the Tb sublattice – a reduction in the value of $m_{Tb}$(x=0) of 23\% for a dose of $1\times10^{17}$ ions/cm$^2$ and a reduction of 26\% for $2\times10^{17}$ ions/cm$^2$. It is important to note that because of the assumptions made for this fitting, this represents the minimum reduction in the Tb sublattice. While we presume that the Co magnetic moment does not change significantly, if the Co sublattice also reduces in magnetic moment then the corresponding reduction in the Tb sublattice must be greater to achieve compensation at the same composition observed here and element sensitive techniques such as XMCD would be required to distinguish this in greater detail.

\begin{figure*}
    \centering
    \includegraphics[width=1.8\columnwidth]{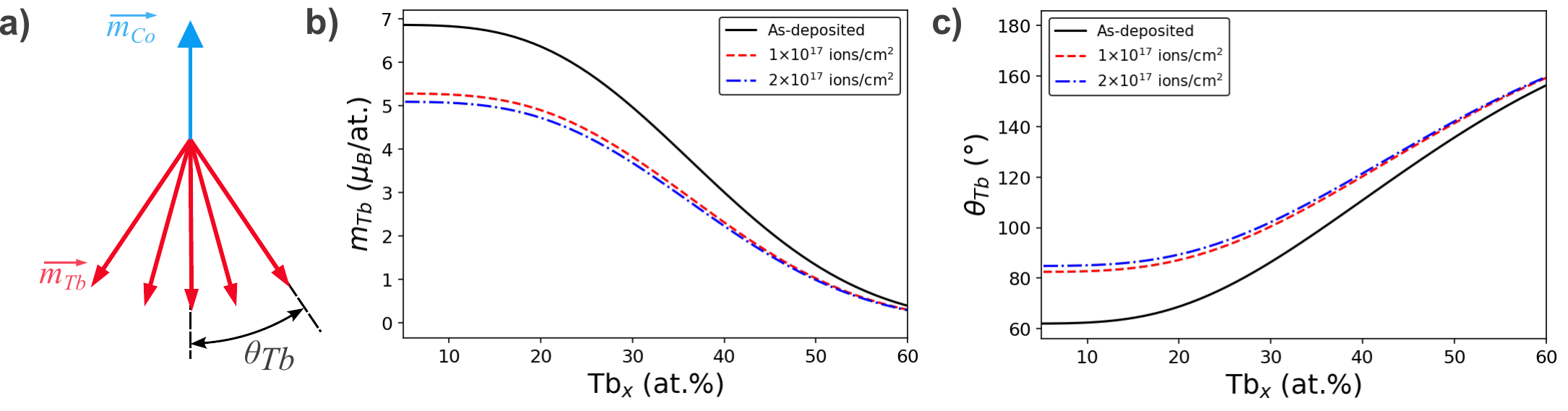}
    \caption{a) Schematic diagram of the sperimagnetic structure in TbCo alloys. b) $m_{Tb}(x)$ profiles with increasing hydrogen dose obtained from the environment model fits in Fig. \ref{fig:ms_comp}. c) Resulting sperimagnetic cone angle assuming any reduction in the magnetic moment results from changes to the sperimagnetic structure.}
    \label{fig:speri}
\end{figure*}

An important element to consider with this material is the sperimagnetic structure - that is, the disordered aspect of the alignment between the Tb and Co sublattices \cite{rhyne1979amorphous}. The sperimagnetic structure arises from the amorphous structure of the film with each site having different local energy minima leading to deviations in the alignment of the magnetic moments which can be described as a fan or cone structure. The Co-Co exchange remains the most dominant factor for Co atoms and so they align parallel and are typically described as having no conical structure \cite{uchiyama1995magnetic}. For the Tb sites, the Tb-Co exchange dominates and local fluctuations in composition have a large impact on the structure giving rise to the sperimagnetic cone structure, illustrated in Fig. \ref{fig:speri}a.

As a result, we can also consider changes in the atomic magnetic moment of Tb as a change in the sperimagnetic structure of the Tb$_x$Co$_{(100-x)}$ film. From the changes in PMA strength, we can understand that the hydrogen is disrupting the random single-ion anisotropy. This random anisotropy is also a leading factor in the existence of the sperimagnetism in these materials and so changes in the Tb atomic magnetic moment could be understood through the lens of changes to the sperimagnetic cone angle, as has been done in Ref. \cite{vas2015specific}. For a filled cone structure the cone angle, $\theta_{Tb}$, can be calculated by comparing the measured atomic magnetic moment, $m_{Tb}$, to the expected moment of a free terbium atom of $\mu_{Tb}=9.34\mu_B$ \cite{makarochkin2020features},

\begin{equation}
    m_{Tb} = \mu_{Tb} (1+\cos{\theta_{Tb}})/2,
\end{equation}

which can be numerically determined based on the compositional profiles of $m_{Tb}$ obtained from the fits to the environment model obtained previously and shown in Fig. \ref{fig:speri}b). The resulting obtained cone angle curves shown in Fig. \ref{fig:speri}c) then follow the inverse of the $m_{Tb}(x)$ curve, with the largest changes at the smallest values of $x$ and the hydrogen dose-dependent curves converging for high values of $x$ at which point the atomic magnetic moment tends towards zero. The reduction in $m_{Tb}$ in this model would imply that as a result of hydrogen uptake there is an increase in the dispersion of the Tb sublattice magnetizations. 

It is however challenging to distinguish a reduction in the measured magnetization due to an increased cone angle from a reduction in the Tb-Co exchange interaction. Measurement of the cone angle requires characterization techniques that are sensitive the orthogonal components of magnetization, such as neutron diffraction \cite{givord1985exchange} or Mössbauer spectroscopy \cite{coey1976magnetic} (although this would mostly be limited to alloys containing Fe). Additionally, further characterization of the local atomic and electronic structural changes are needed to understand the change in moment fully.

\section{Conclusions}

In conclusion we have evaluated the effect of hydrogen ion implantation with different doses on the magnetic properties of magnetic properties of amorphous Tb$_x$Co$_{(100-x)}$ alloys for a range of alloy compositions. The effect of hydrogen loading in this way can clearly be seen as an effective shift in the compensation composition and in the Co-rich in-plane to out-of-plane transition. From our magnetometry, the change in compensation composition can be linked to a reduction in the magnetic moment of the Tb sublattice primarily. 

The origin of these effects is likely a result of the change in metal pair correlations increasing in length due to the implanted hydrogen, but whether this results in a reduction of the exchange coupling, a change in the sperimagnetic structure, or both requires further experiments to determine. 

These results will be important for future work focusing on devices aiming to take advantage of recent advances in solid-state proton-pump methods by providing quantitative information that relates the amount of hydrogen loaded with the change in magnetic properties, as well as providing insight into ideal choices of composition. 

\section{Dava Availability}

Data is available from the corresponding author upon reasonable request.

\begin{acknowledgments}

The authors would like to thank Johan Oscarsson and Mauricio
Sortica at the Uppsala Tandem Laboratory for their help with ion implantations. R. G. H. acknowledges the support of a Carl-Trygger Foundation Postdoctoral Fellowship (CTS 22:2039, grant holder G.A.). Accelerator operation at the Ion Technology Center national infrastructure is supported by the Swedish Research Council VR-RFI (grant \# 2019$\_$00191).

\end{acknowledgments}


\bibliography{references}

\end{document}